\documentclass[preprint]{ptephy_v1}

\preprintnumber{OU-HET-1135} 

\usepackage{comment}
\usepackage{physics}
\usepackage{mathrsfs}
\usepackage{braket}
\makeatletter
\def\lastpage@putlabel{}
\makeatother
\usepackage{lastpage,hyperref}
\usepackage{graphics} 
\usepackage{breakurl}
\usepackage{subfigure}
\usepackage[utf8]{inputenc}

\newcommand{\fix}[1]{\textcolor{red}{[#1]}}

\newcommand{\Slash}[1]{{\ooalign{\hfil/\hfil\crcr\(#1\)}}}

\begin{document}

\title{Curved domain-wall fermions}


\author{Shoto Aoki}
\author{Hidenori Fukaya}
\affil{Department of Physics, Osaka University,\\Toyonaka, Osaka 560-0043, Japan \email{saoki@het.phys.sci.osaka-u.ac.jp}\email{hfukaya@het.phys.sci.osaka-u.ac.jp}}






\begin{abstract}
We consider fermion systems on a square lattice with a mass term having a curved domain-wall. Similarly to the conventional flat domain-wall fermions, massless and chiral edge states appear on the wall. In the cases of $S^1$ and $S^2$ domain-walls embedded into flat hypercubic lattices, we find that these edge modes feel gravity through the induced Spin or Spin$^c$ connections. The gravitational effect is encoded in the Dirac eigenvalue spectrum as a gap from zero. In the standard continuum extrapolation of the square lattice, we find a good agreement with the analytic prediction in the continuum theory. We also find that the rotational symmetry of the edge modes is automatically recovered in the continuum limit.

\end{abstract}

\subjectindex{xxxx, xxx}

\maketitle

\section{Introduction}


Lattice gauge theory provides a nonperturbative regularization of quantum field theory. In the standard formulation of quantum chromo-dynamics (QCD), we consider a four-dimensional square lattice with periodic boundary conditions, whose continuum limit is supposed to be a flat torus. Since the degrees of freedom become finite in this regularization, the QCD partition function is expressed by a mathematically well-defined integral, which allows a nonperturbative first-principle computation of hadronic processes using numerical simulations.

In contrast to the remarkable success of lattice QCD on a flat Euclidean spacetime, it is not straightforward to formulate a lattice field theory with a gravitational background. Unlike the standard gauge field which is  just put as a link variable on a fixed square lattice, the nontrivial metric or vielbein needs some deformation of the lattice itself. In order to systematically achieve such a deformation, previous works \cite{Hamber2009Quantum,Regge1961general,brower2016quantum,AMBJORN2001347Dynamicallytriangulating,Brower2017LatticeDirac,Catterall2018Topological,ambjorn2022topology} employed triangular lattices, which correspond to the triangulation of manifolds known in mathematics. They tried to represent dynamical or non-dynamical gravity by changing the lengths and/or angles of the link variables. 

One of the problems in such triangular lattice approaches is an ambiguity in taking the continuum limit. On the standard flat square lattice, only one parameter: lattice spacing $a$ (or equivalently gauge coupling) is enough to tune for approaching the continuum limit. It is known that the rotational symmetry (or Lorentz invariance after the Wick rotation) is automatically recovered in that limit if the lattice action respects a discrete subgroup symmetry of it. On the other hand, there is no such simple and unique way for making a finer triangular lattice from a given triangular lattice. It was reported in \cite{Brower2015Quantum} that some counter terms are required to recover the rotational symmetry of spherical manifolds.

In this work, we attempt to formulate fermion systems with a nontrivial gravitational background put on a square lattice. In mathematics, it is known that any Riemannian manifold can be isometrically embedded into a sufficiently higher-dimensional Euclidean space \cite{Nash1956TheImbedding,Gromov1970Embeddingsand}. If we regularize this higher dimensional flat space by a square lattice and localize the fermion field on the embedded submanifold, the fermion would feel gravity through the Spin connection induced by the embedding. Since the total system is given by a flat square lattice, the continuum limit is naturally taken by reducing the lattice spacing, just in the same way as the standard lattice gauge theory. If the action respects the symmetry under right-angle rotations of the whole system, it is also natural to assume that the rotational symmetry, as well as that of the embedded manifold if exists, will be automatically recovered in this simple continuum limit.

It is well-known in the so-called domain-wall fermion formulation \cite{Jackiw1976Solitons,CALLAN1985427Anomalies,KAPLAN1992342AMethod,Shamir1993Chiral,Furman1995Axial} that edge-localized states appear on the codimension-one subspace. Moreover, these modes are massless and chiral when the domain-wall is even dimensions. In lattice QCD, the edge-localized modes are regarded as quarks, and their effective Dirac operator satisfies the Ginsparg-Wilson relation \cite{Ginsparg1981ARemnant} in the infinite limit of the extra dimension, with which one has an exact chiral symmetry on a lattice \cite{Luscher1998Exactchiral}. With the Dirac operator satisfying the Ginsparg-Wilson relation, the Atiyah-Singer index can be defined even with finite lattice spacings \cite{HASENFRATZ1998125Theindex}. Recently, the Atiyah-Patodi-Singer index on a manifold with boundaries was reformulated using the domain-wall fermions in continuum theory \cite{Fukaya_2017Atiyah-Patodi-Singer,fukayaFuruta2020physicistfriendly,FukayaFurutaMatsuki2021Aphysicist-friendly}, which was extended to the lattice gauge theory \cite{FukayaKawai2020TheAPS}. However, as far as we know, the previous works were limited to the cases where the domain-wall is a flat Euclidean space.



In the continuum theory, localized states at the curved subspace have been actively studied. Historically, a free non-relativistic system bounded on a submanifold was first considered in \cite{JENSEN1971586Quantummechanics,daCosta1982Quantummechanics} and then they extended the work to the case with an external gauge field in \cite{Pershin2005Persistent,Ferrari2008Schrodinger}. Furthermore, a similar study was done for the relativistic Dirac field in \cite{BRANDT20163036Diracequation,Matsutani1992Physicalrelation,Matsutani1994TheRelation,Matsutani1997Aconstant,Burgess1993Fermions}. These studies pointed out that a geometric potential or a nontrivial Spin connection is induced in the effective theory on the surface, and such potentials were experimentally discovered in \cite{Szameit2010GeometricPotential,Onoe2012ObservationofRiemannian}. Topological insulators having a curved surface or bubbles inside were also considered. In \cite{Lee2009Surface,Imura2012Spherical,Parente2011Spin,Takane2013UnifiedDescription}, they found that the edge-localized modes appearing on the spherical surface feel gravity through the non-trivial connection. In a relativistic framework, the curved domain-wall was studied in the context of anomaly inflow \cite{CALLAN1985427Anomalies}. For example, it was shown in \cite{catterall2022induced}, that there exists a gravitational anomaly \cite{Butt2021Anomalies,Catterall2018Topological} in a three-dimensional K\"{a}hler-Dirac fermion system and its anomaly is canceled by a contribution from the curved domain-wall.

In this work, we consider embedding of a circle $S^1$ as a domain-wall into a two-dimensional Euclidean flat lattice, as well as of a sphere $S^2$ domain-wall put on a three-dimensional square lattice. On each domain-wall, we find massless states localized at the curved domain-wall, and the Dirac eigenvalue spectrum shows a gravitational effect through the induced Spin connection on the wall. A preliminary result has been already presented in \cite{aoki2021chiral}.

The rest of the paper is organized as follows. In Sec.~\ref{sec:Aconnection}, we propose a fermion system with a general curved domain-wall mass on a general Spin Riemannian manifold. We discuss in continuum theory how the edge modes emerge at the domain-wall and how the induced Spin connection is detected from the Dirac operator. In Sec.~\ref{sec:S^1 in R^2} and Sec.~\ref{sec:S^2 in R^3}, we explicitly solve the eigenproblem of the edge modes in the two-dimensional system with an $S^1$ domain-wall, as well as an $S^2$ embedded in three dimensions. We compare the numerical lattice results and those in continuum and estimate the finite volume corrections. A special focus is placed on the recovery of the rotational symmetry. Finally, we give a summary and discussion in Sec.~\ref{sec:Conclusion}.

\section{Curved manifold embedded into Euclidean space}\label{sec:Aconnection}
It was shown by Nash \cite{Nash1956TheImbedding,Gromov1970Embeddingsand} that any Riemannian manifold can be isometrically embedded into a sufficiently higher dimensional flat Euclidean space. Conversely, if we have an embedding function of a manifold into a Euclidean space its metric is uniquely determined (induced) by the embedding. In this section, we compute the induced metric as well as the associated Spin connection by the embedding. We also discuss how the edge-localized modes of fermion emerge when the embedding is given by a domain-wall mass term.

\subsection{Induced Spin connection by embedding}


Let us consider an $n$-dimensional Riemannian manifold $Y$ with a metric $h$, which is isometrically embedded into a Euclidean space $\mathbb{R}^m$ with the $m$-dimensional flat metric $\delta$ (the integer $m$ can be taken finite in a range $m \leq \frac{1}{2}(n + 2)(n + 5)$ \cite{Nash1956TheImbedding,Gromov1970Embeddingsand}). Let us denote the embedding function by $x^I(y^1,y^2,\cdots y^n)$ where $x^{I=1,2,\cdots m}$ denotes the coordinate on $X=\mathbb{R}^m$ as a function of the coordinate $y^{i=1,2,\cdots n}$  on $Y$. Then, the induced metric $h$ is uniquely determined (up to the diffeomorphism on $Y$) by 
\begin{align}
    h_{ab}=\sum_{IJ} \delta_{IJ} \pdv{x^I}{ y^{a}} \pdv{x^J}{ y^{b}}.
\end{align}

If a particle is constrained on the curved manifold $Y$, it feels gravity by the equivalence principle, through the induced metric $h$. In continuum theory, this was confirmed in \cite{JENSEN1971586Quantummechanics,daCosta1982Quantummechanics,Pershin2005Persistent} and experimentally realized in \cite{Szameit2010GeometricPotential,Onoe2012ObservationofRiemannian}. What about relativistic Dirac fermion fields? According to \cite{BRANDT20163036Diracequation,Matsutani1992Physicalrelation,Matsutani1994TheRelation,Matsutani1997Aconstant,Burgess1993Fermions}, if $Y$ is a Spin manifold, its Spin connection is also induced by the embedding.

Let us consider a tangent vector space at $p \in Y$. Since $p$ is also a point of $X(=\mathbb{R}^m)$, we can decompose the tangent space $T_p X$ as 
\begin{align}
    T_p X \simeq T_p Y \oplus N_p,
\end{align}
where $N_p$ is a normal vector space to $T_p Y$. Let $\qty{e_1,\cdots,e_{n}}$ be a vielbein of $Y$, then we can choose a vielbein of $X$ as
\begin{align}
    \{\underbrace{e_1,\cdots,  e_n}_{\text{vielbein of $Y$ }} ,\underbrace{e_{n+1} ,\cdots ,e_m}_{\text{normal vector}}\}, \label{eq:vielbein}
\end{align}
where $e_I=e_I^{\ J} \pdv{}{x^J}$. The component $e_I^{~J}$ is determined as the following. On $Y$, the vector $\pdv{}{y^a}$ is written as
\begin{align}
    \pdv{}{y^a}=\sum_{I=1}^m \pdv{x^I}{y^a} \pdv{}{x^I}.
\end{align}
We can regard  $\qty( \pdv{x^I}{y^a})$ as a $m\times n $ matrix with the rank $n$ to obtain the basis transformation
\begin{align}
    &\qty( \pdv{}{y^1}, \cdots, \pdv{}{y^n}, \pdv{}{x^{n+1}}, \cdots,\pdv{}{x^m})\nonumber \\
    &= \qty( \pdv{}{x^1}, \cdots, \pdv{}{x^n},\pdv{}{x^{n+1}},  \cdots, \pdv{}{x^m})\left( \begin{array}{ccc|ccc}
\pdv{x^1}{y^1} & \cdots & \pdv{x^1}{y^n} &  &  &  \\
\vdots & \ddots & \vdots &  &  &  \\
\pdv{x^n}{y^1} & \cdots & \pdv{x^n}{y^n} &  &  &  \\ \hline
\pdv{x^{n+1}}{y^{1}} & \cdots & \pdv{x^{n+1}}{y^{n}} & 1 &  &  \\
\vdots & \ddots & \vdots &  & \ddots &  \\
\pdv{x^{m}}{y^1} & \cdots & \pdv{x^{m}}{y^n} &  &  & 1 \\
    \end{array} \right).
\end{align}
The vielbein can be obtained by the Gram–Schmidt orthonormalization:
\begin{align}
\begin{aligned}
    e_1&= \frac{e_1^\prime}{\norm{e_1^\prime}} ,~\qty(e_1^\prime= \pdv{}{y^1}) \\
    e_2&=\frac{e_2^\prime}{\norm{e_2^\prime}},~\qty(e_2^\prime= \pdv{}{y^2} -\delta\qty(e_1,\pdv{}{y^2}) e_1) \\
    e_3&= \cdots,
\end{aligned}
\end{align}
where $\norm{v}=\sqrt{\delta(v,v)}$ for vector fields $v$ on $X$. 

Although $X$ is a flat Euclidean space, the above choice of the vielbein makes the Levi-Civita and Spin connections look nontrivial. From the torsionless condition and vielbein postulate, we uniquely obtain the Christoffel symbol
\begin{align}
    \Gamma^I_{JK}=\frac{1}{2} g^{I A}\qty{ \pdv{g_{AJ}}{x^K }+\pdv{g_{AK}}{x^J } - \pdv{g_{JK}}{x^A}},~g^{IJ}=\sum_{K} e_K^{~I}e_K^{~J},
\end{align}
as well as the Spin connection (when $Y$ is a Spin manifold)
\begin{align}
    \omega_K= \frac{1}{4}\sum_{KJ}\omega_{IJ,K} \gamma^I \gamma^J ,~{\omega}^I_{\ J,K}=-\frac{1}{2}\qty( C^I_{J,K}+C^J_{K,I} -C^K_{I,J}),
    \label{eq:connection expressed by a frame}
\end{align}
where $C^{K}_{I,J}$ is defined by the commutator $[e_I,e_J]=e_K C^K_{I,J}$ and written as
\begin{align}
     C^{K}_{I,J}=g^{NM} e^{\ N}_K \qty( e_I^L \pdv{e_J^{\ M}}{x^L} - e_J^L \pdv{e_I^{\ M}}{x^L} ).
\end{align}

The Spin connection on $Y$ can be identified by simply
collecting those having indices in $\{1,\cdots n\}$:
\begin{align}
    \omega_c = \frac{1}{4}\sum {\omega}_{ab,c} \gamma^a \gamma^b.
\end{align}
Note that the above connections give zero curvature everywhere on $X$.

Next, we consider a domain-wall fermion system where the mass term flips its sign on a codimension one manifold $Y$ embedded into $X=\mathbb{R}^{n+1}$. Let $f:X \to \mathbb{R}$ be a smooth function such that $f^{-1}(0) \neq \emptyset$ and $df|_{p} \neq 0$ for all $p\in f^{-1}(0)$. Then $Y:=f^{-1}(0)$ is a hypersurface in $X$ and an $n$-dimensional smooth manifold by Preimage theorem. $Y$ can be regarded as a domain-wall dividing $X$ into the $f>0$ and $f<0$ regions. We can take a vielbein $\qty{e_1,\cdots ,e_{n+1}}$ as \eqref{eq:vielbein}. Since $\qty{e_1,\cdots ,e_n}$ are tangent vectors of $Y$,
\begin{align}
    e_a(f)=e_a^I\pdv{f}{x^I}=0
\end{align}
or equivalently we have
\begin{align}
    e_{n+1}=\frac{1}{\norm{\text{grad}(f)}} \text{grad}(f), 
\end{align}
where $ \text{grad}(f)=\sum_{I} g^{IJ}\pdv{f}{x^I} \pdv{}{x^J}$. The Dirac operator on $X$ is
\begin{align}
\begin{aligned}
    \Slash{D}+m\epsilon
    =\gamma^a\qty(e_a+\frac{1}{4}\sum_{bc}\omega_{bc,a} \gamma^b\gamma^c+ \frac{1}{2}\sum_{b}\omega_{b\ n+1,a}\gamma^b \gamma^{n+1}) \\
   +\gamma^{n+1}\qty(e_{n+1}+\frac{1}{4}\sum_{bc}\omega_{bc,n+1} \gamma^b\gamma^c+\frac{1}{2} \sum_{b}\omega_{b\ n+1,n+1} \gamma^b \gamma^{n+1})+m\epsilon,
\end{aligned}
\end{align}
where $\epsilon=\text{sign}(f)$ is a step function.

Let us decompose the above Dirac operator into the one on $Y$ and that in the normal direction. First, note that we can take $\omega^b_{~c,n+1}=\omega_{bc,n+1}=0$ by a local $Spin(n)$ rotation. Let us denote the coordinate in the normal direction by $t$. Then $\omega_{bc,n+1}$ is absent in the transformed Dirac operator
\begin{align}
\Slash{D} &\to L^{-1}(y,t)\Slash{D}L(y,t),\\
L(y,t) &= P\exp\left[-\frac{1}{4}\sum_{bc}\gamma^b\gamma^c
\int_0^t dt'\omega_{bc,n+1}(y,t')  \right],
\end{align} 
where $P$ denotes the path-ordered product.

Next, we compute $\omega^{b}_{\ n+1,n+1}=\omega_{b\ n+1,n+1}$. According to \eqref{eq:connection expressed by a frame}, it is enough to obtain $C^{n+1}_{b,n+1}$. Using the commutator
\begin{align}
    [e_b,e_{n+1}]&=C^c_{b,n+1}e_c+C_{b,n+1}^{n+1}e_{n+1},
\end{align}
and $df(e_{n+1})=\delta(\text{grad}(f),e_{n+1})= \norm{\text{grad}(f)}$, we have a relation
\begin{align}
    0&=ddf(e_b,e_{n+1})=e_b df(e_{n+1})-e_{n+1} df(e_b)-df([e_b,e_{n+1}]) \nonumber \\
    &=e_b \qty(  \norm{\text{grad}(f)}) -C^{n+1}_{b,n+1} \norm{\text{grad}(f)}.
\end{align}  
Since $\norm{\text{grad}(f)}$ is nonzero around $Y$, we obtain 
\begin{align}
    \omega^{b}_{\ n+1,n+1}&=-\frac{1}{2} \qty(C^{b}_{n+1,n+1}+C^{n+1}_{n+1,b}-C^{n+1}_{b,n+1})=C^{n+1}_{b,n+1} \nonumber \\
   &=\frac{1}{\norm{\text{grad}(f)}}e_b \qty(  \norm{\text{grad}(f)})=
    \frac{1}{2}e_b\qty( \log( g^{IJ} \pdv{f}{x^I} \pdv{f}{x^J})).
\end{align}

Finally, let us denote the remaining connection $\omega^b_{~n+1,a}$ by $\omega^b_{~n+1,a}=-h_{ab}$, which is known as the second fundamental form or shape operator. 
Since $\omega^a_{~n+1,b}-\omega^{b}_{~n+1,a}=C^{n+1}_{a,b}=0$, $h_{ab}$ is symmetric: $h_{ab}=h_{ba}$. Now the Dirac operator becomes
\begin{align}
    &\Slash{D} +m\epsilon \nonumber \\
    =&\gamma^a\qty(\tilde{\nabla}_{e_a} )+\frac{1}{2}\sum_{ab}(-h_{ab})\gamma^a \gamma^b \gamma^{n+1}\nonumber \\
    &+\gamma^{n+1} \qty(e_{n+1}+\frac{1}{4}\sum_b e_b\qty( \log( g^{IJ} \pdv{f}{x^I} \pdv{f}{x^J}))\gamma^b \gamma^{n+1}) +m\epsilon \nonumber\\
    =&\gamma^a\qty(\tilde{\nabla}_{e_a}) +\gamma^{n+1}\qty(e_{n+1} -\frac{1}{2}\tr h +\frac{1}{4} e_a\qty( \log( g^{IJ} \pdv{f}{x^I} \pdv{f}{x^J}) ) \gamma^a \gamma^{n+1}) +m\epsilon.
\end{align}
Here $\frac{1}{n}\tr h=\frac{1}{n} \sum_a h_{aa}$ corresponds to the
mean curvature of the $n$-dimensional Riemannian submanifold $Y$. This expression is consistent with the previous works \cite{BRANDT20163036Diracequation,Matsutani1992Physicalrelation, Matsutani1994TheRelation,Matsutani1997Aconstant,Burgess1993Fermions}. 

In order to solve the Dirac equation, it may be convenient to perform a rescaling transformation $\psi= \qty(g^{IJ}\pdv{f}{x^I}\pdv{f}{x^J})^{\frac{1}{4}}\psi^\prime$. On $\psi^\prime$, the Dirac operator acts as
\begin{align}
\begin{aligned}
    \Slash{D}^\prime +m\epsilon=& \gamma^a \qty(e_a +\frac{1}{4}\sum_{bc}\omega_{bc,a} \gamma^b\gamma^c)  \\
    &+\gamma^{n+1}\qty( e_{n+1} -\frac{1}{2}\tr h + \frac{1}{4} e_{n+1} \qty(\log( g^{IJ}\pdv{f}{x^I}\pdv{f}{x^J}))) +m\epsilon.
\end{aligned}
\end{align}
The first term corresponds to the massless Dirac operator on $Y$, where we can find the induced gravity effect as a nontrivial Spin connection. This form suggests the existence of massless edge modes no matter how $Y$ is curved, when we find a function of the normal coordinate $t$ on which the second and third terms of the operator give zero.


As a final remark of this subsection, we note that the above result for the Dirac operator is valid only locally, although the original coordinates and operator are defined globally on $X=\mathbb{R}^m$. To obtain the global solution of the Dirac equation, we need to separately find the solutions on the open patches covering $X$ and check the consistency conditions on them. As will be discussed below, the Berry phase of the spinor field may help to simplify this issue. We also note that when $Y$ is a closed manifold the Spin structure of $Y$ must belong to the trivial element of the Spin bordism group. 


\subsection{Edge-localized modes on the domain-walls}\label{subsec:edge modes}


In this subsection, let us examine the existence of the edge-localized modes formally solving the Dirac equation. Let $Y$ be an $n$-dimensional submanifold of $X=\mathbb{R}^{n+1}$ embedded by a function $f: X\to \mathbb{R}$. We can express the Dirac's gamma matrices by
\begin{align}
    \gamma^a=-\sigma_2 \otimes \tilde{\gamma}^a,\ \gamma^{n+1}=\sigma_1 \otimes 1 ,\ \bar{\gamma}=\sigma_3 \otimes 1,  
\end{align}
where $\tilde{\gamma}^a~(a=1,\cdots n)$ are the $2^{[n/2]}\times 2^{[n/2]}$ gamma matrices ($[\alpha]$ denotes the Gauss symbol or the integer part of $\alpha$) which satisfy the Clifford algebra in $n$ dimensions, and $\sigma_{1,2,3}$ are the Pauli matrices. When $n$ is even, the above gamma matrices are not in the irreducible representation and the Pauli matrices can be interpreted as the operators on the two-flavor space. In general, the massive Dirac operator is a complex operator in odd dimensions. In order to make a well-defined eigenvalue problem of a Hermitian operator, we introduce these two flavors of the spinor fields\footnote{In the appendix~\ref{App:Weyl} we consider the one-flavor ``chiral'' case where the Dirac operator is non-Hermitian.}. Since $\bar{\gamma}$ anti-commutes with any other $\gamma$'s, we call it the ``chirality'' operator on $X$. Unlike the standard flat domain-wall fermion, however, the edge-localized modes are eigenstates of another ``chirality'' operator, different from $\bar{\gamma}$, as is discussed below.

Expressing the Dirac spinor on $X$ by
\begin{align}
    \psi=\mqty( \chi_1 \\ \chi_2),
\end{align}
let us solve the eigenplobrem of the Hermitian Dirac operator on $\psi$,
\begin{align}
    H=\bar{\gamma}(\Slash{D}+\epsilon m). \label{eq:Hermition Dirac operator for odd in even}
\end{align}

Around $p\in Y$, we can set a chart $(y^1,\cdots,y^n ,t)$ and its associated frame, where $(y^1,\cdots ,y^n)$ denotes the coordinate along $Y$ and $t$ is that in the direction of the normal vector $e_{n+1}$. Here $t=0$ denotes where $Y$ is located.
On this chart, $\epsilon=\text{sign}(t)$. Let $\psi= \qty(g^{IJ}\pdv{f}{x^I}\pdv{f}{x^J})^{\frac{1}{4}}\psi^\prime$. On $\psi^\prime$, the Dirac operator acts as
\begin{align}
    H^\prime
    =&\bar{\gamma}(\Slash{D}^\prime+\epsilon m)=\mqty(\epsilon m& i\tilde{\Slash{D}}+\pdv{}{t}+F \\i\tilde{\Slash{D}}-\pdv{}{t}-F & -\epsilon m ),
\end{align}
where $F=-\frac{1}{2}\tr h + \frac{1}{4} \pdv{}{t} \qty(\log( g^{IJ} \pdv{f}{x^I} \pdv{f}{x^J}))$ is a function of $(y^1,\cdots ,y^n,t)$ in general and $i\tilde{\Slash{D}}=\tilde{\gamma}^a \tilde{\nabla}_a$ is the Dirac operator on $Y$.

Let us consider a solution with an eigenvalue much smaller than $m$.
In the large $m$ limit, the following part of the operator $H'$ must vanish separately.
\begin{align}
H'_{\rm normal} &=i\sigma_2\qty(\pdv{}{t} +F+\sigma_1 m \epsilon)\otimes 1, 
\end{align}
which requires the solution to have the form
\begin{align}
    \psi'= e^{-m \abs{t}}\left[\exp(-\int_0^t dt^\prime F(y,t^\prime)) \mqty(\chi(y) \\ \chi(y))+\order{t}\right].
\end{align}
If we take $\chi(y)$ to be an eigenstate of $i\tilde{\Slash{D}}|_{t=0}$ with the eigenvalue $\lambda$,
we obtain the edge-localized mode of $H$ in the $m\to \infty$ limit by
\begin{align}
    \psi= \qty(g^{IJ}\pdv{f}{x^I}\pdv{f}{x^J})^{\frac{1}{4}}e^{-m \abs{t}}
    \exp(-\int_0^t dt^\prime F(y,t^\prime)) \mqty(\chi(y) \\ \chi(y)),
\end{align}
which has the negative eigenvalue of the ``chirality'' operator
$\gamma^{n+1}=\sigma_1\otimes 1$ and the eigenvalue $\lambda$ of the
massless $n$-dimensional operator: $i\tilde{\Slash{D}}|_{t=0}\chi(y)=\lambda\chi(y)$.
These modes feel gravity through the induced Spin connection in the effective Dirac operator on $Y$.

Finally, let us evaluate the leading order contribution of the finite $1/m$ corrections.
From the above solution, we have
\begin{align}
    (H-\lambda) \psi= \qty(g^{IJ}\pdv{f}{x^I}\pdv{f}{x^J})^{\frac{1}{4}} e^{-m\abs{t}}e^{-\int_0^t dt^\prime F(y,t^\prime)} \underbrace{\qty(-\int_0^t dt^\prime \gamma^a e_a( F(y,t^\prime)) )}_{\order{t}} \mqty( \chi(y) \\ \chi(y) ).
\end{align}
In the  limit of $m\gg \abs{F(y,t)}$, we can estimate the magnitude of the residual error as
\begin{align}
    \norm{(H-\lambda)\psi} \leq \frac{C}{m},
\end{align}
where $\norm{\ast}$ denotes the norm of the Dirac spinor on $X$ and $C$ is a positive number independent of $m$.
Therefore, $\psi$ becomes an eigenstate of $H$ when $m$ is large enough.

\section{$S^1$ domain-wall in two-dimensional flat space}\label{sec:S^1 in R^2}

In this section, we embed a one-dimensional sphere $S^1$ as a domain-wall into the flat two-dimensional space. This situation is the simplest case to which the argument of Sec.~\ref{subsec:edge modes} applies.

\subsection{Continuum analysis}
 The Hermitian Dirac operator \eqref{eq:Hermition Dirac operator for odd in even} in this case is 
\begin{align}
    H&=\sigma_3 \qty(\sigma_1 \pdv{}{x} +\sigma_2 \pdv{}{y} + m\epsilon) 
    =\mqty(m\epsilon & e^{-i\theta}(\pdv{}{r}-\frac{i}{r}\pdv{}{\theta}) \\
    -e^{i\theta}(\pdv{}{r}+\frac{i}{r}\pdv{}{\theta})& -m\epsilon
    ) \label{eq:Dirac op S^1 conti},
\end{align}
in the continuum Euclidean space $\mathbb{R}^2$, where an $S^1$ domain-wall with radius $r_0$ is put by the sign function $\epsilon= \text{sign}\qty( r-r_0)$. We take the polar coordinates $(r,\theta)$ so that the radial direction is equal to the normal direction to the $S^2$ domain-wall. Here $\psi$ is a two-component spinor on $\mathbb{R}^2$.

In \eqref{eq:Dirac op S^1 conti}, the sigma matrix in the normal direction and that in the tangent direction are
\begin{align}
\sigma_r = \sigma_1 \cos\theta +\sigma_2 \sin \theta,\ 
\sigma_\theta = \sigma_2\cos\theta -\sigma_1 \sin\theta.
\end{align}
By a local $Spin(2)$ gauge transformation, we can take a frame where they are universally given by $\sigma_r \to e^{-i\frac{\theta}{2}\sigma_3}\sigma_r e^{i\frac{\theta}{2} \sigma_3}=\sigma_1$ and $\sigma_\theta \to e^{-i\frac{\theta}{2}\sigma_3}\sigma_\theta e^{i\frac{\theta}{2}\sigma_3}=\sigma_2$, respectively. However, this transformation spoils the original periodic property of the spinor $\psi(r,\theta+2\pi)=\psi(r,\theta)$.
As discussed in the previous section, this simply reflects the fact that the frame can be taken only locally and a careful gluing of them is required to give a consistent result.
Here we avoid this issue by a further $U(1)$ transformation
$e^{-i\frac{\theta}{2}}$, which produces a Berry connection (or $Spin^c(1)$ connection in mathematics).
Specifically, we transform the spinor as $\psi \to \psi^\prime=e^{i\frac{\theta}{2} \sigma_3} e^{-i\frac{\theta}{2}}\psi$ keeping its periodicity to obtain the transformed Dirac operator
\begin{align}
    H^{\prime}=&e^{i\frac{\theta}{2} \sigma_3} e^{-i\frac{\theta}{2}} H e^{-i\frac{\theta}{2} \sigma_3} e^{+i\frac{\theta}{2}} \nonumber\\
    =& \sigma_3 \qty(\sigma_1 \qty(\pdv{}{r}+\frac{1}{2r})  +\sigma_2 \frac{1}{r} \qty(\pdv{}{\theta} +i\frac{1}{2})+ m\epsilon)
\end{align}
which operates on the periodic spinor.

The explicit form of the eigenfunction with the the eigenvalue $-m<E<m$ of $H^\prime$ is given by
\begin{align}
    (\psi^{E,j})^{\prime}&=\left\{
\begin{array}{ll}
A \mqty(\sqrt{m^2-E^2} I_{j-\frac{1}{2}} (\sqrt{m^2-E^2} r)e^{i(j-\frac{1}{2})\theta }\\ 
         (m+E) I_{j+\frac{1}{2}} (\sqrt{m^2-E^2} r)e^{i(j-\frac{1}{2})\theta }) & (r<r_0) \\
B\mqty((m+E)K_{j-\frac{1}{2}} (\sqrt{m^2-E^2} r)e^{i(j-\frac{1}{2})\theta }\\ 
         \sqrt{m^2-E^2} K_{j+\frac{1}{2}} (\sqrt{m^2-E^2} r)e^{i(j-\frac{1}{2})\theta }) & (r>r_0)
\end{array}
\right. ,
\end{align}
where $j$ takes a half-integer value $j=\pm \frac{1}{2},\pm \frac{3}{2},\cdots$. Here $I_n$ and $K_n$ are the modified Bessel functions. From their exponentially decaying asymptotic form, this eigenfunction represents an edge-localized mode at $r=r_0$. From the continuity at $r=r_0$, the coefficients $A$ and $B$ are determined (with the normalization condition) and we obtain a nontrivial eigenvalue condition 
\begin{align}\label{eq:condition of E}
    \frac{I_{j-\frac{1}{2}}}{I_{j+\frac{1}{2}}}\frac{K_{j+\frac{1}{2}}}{K_{j-\frac{1}{2}}}(\sqrt{m^2-E^2}r_0)=\frac{m+E}{m-E}.
\end{align}

In the large mass limit or $m\gg E$, the eigenvalue converges to
\begin{align}
    E\simeq \frac{j}{r_0}\ \qty(j=\pm\frac{1}{2},\pm\frac{3}{2},\cdots) \label{eq:S1 eigenavlue}. 
\end{align}
The normalized eigenfunction in that limit is also simplified as\footnote{
This approximation is only valid at $r\sim r_0$.
}
\begin{align}\label{eq:asymptotic form of edgemode S^1}
    (\psi_\text{edge}^{E,j} )^\prime\simeq \sqrt{\frac{m}{4\pi r}}e^{-m\abs{r-r_0}}\mqty( e^{i(j-\frac{1}{2})\theta}\\ e^{i(j-\frac{1}{2})\theta}), 
\end{align}
which is chiral with respect to a gamma matrix $\sigma_1$ facing the normal direction to the domain-wall with the eigenvalue $+1$. $\sigma_1$ corresponds to $\sigma_r$ in the original frame of $\mathbb{R}^2$. 

In order to identify the induced gravity effect, let us solve the eigenproblem again taking the $m\to \infty$ limit first. To obtain the finite eigenvalue in this limit, the solution must have the form
\begin{align}
    \psi_\text{edge}^\prime &= \sqrt{\frac{m}{4\pi r}}e^{-m\abs{r-r_0}}\mqty(
1 \\1) \chi^\prime(\theta),
\end{align}
to cancel the $m$ dependence of the operator.
On this edge mode, the Dirac operator effectively acts as 
\begin{align}
    -i\frac{1}{r_0} \qty(\pdv{}{\theta}+i\frac{1}{2}),
\end{align}
where we have used $\sigma_1=1$.

Here, the second term $1/2r_0$ can be identified as the induced Spin$^c$ connection. It is now obvious from the Fourier transformation of $\chi'(\theta)\to \exp(in\theta), n\in \mathbb{Z}$, that the eigenvalue takes a half integer $j=n+1/2$ multiplied by $1/r_0$. The gap in the eigenvalue spectrum seen in \eqref{eq:S1 eigenavlue} opens due to this gravitational effect.

One can locally cancel the gravitational effect or the connection term by a $U(1)$ transformation $\chi^\prime \to \chi=\exp(i\theta/2) \chi^\prime$. However, the global effect of the curved $S^1$ remains as the anti-periodic boundary condition of $\chi$, leaving the eigenvalue spectrum including the gap unchanged. In mathematics, the anti-periodic $\chi$ field is more natural on $S^1$ in the sense that it belongs to a trivial element in the Spin bordism group. The induced Spin$^c$ connection naturally describes how the anti-periodicity emerges in the effective low-energy theory of edge modes, of which the total original system is made periodic.

\subsection{Lattice analysis}
Let us discretize the Dirac operator  \eqref{eq:Dirac op S^1 conti} on a two-dimensional square lattice $(\mathbb{Z}/N\mathbb{Z})^2$. When we take the lattice size $L=1$, the lattice spacing is simply given by $a=1/N$. We represent $\hat{x}=x/a$ and $\hat{y}=y/a$ as the coordinates of $(\mathbb{Z}/N\mathbb{Z})^2$ in the range $0\leq \hat{x},\hat{y} \leq N-1$ and we assume the periodic boundary condition $\hat{x}=0\sim N,\ \hat{y}=0\sim N$. Denoting the difference operators by $(\nabla_1
\psi)_{(\hat{x},\hat{y})}=\psi_{(\hat{x}+1,\hat{y})}-\psi_{(\hat{x},\hat{y})}$
and $(\nabla_1^\dagger
\psi)_{(\hat{x},\hat{y})}=\psi_{(\hat{x}-1,\hat{y})}-\psi_{(\hat{x},\hat{y})}$
and those in the $\hat{y}$ in the same manner,
we have the lattice version of the domain-wall Dirac operator
\begin{align}\label{eq:Hermitian Wilson Dirac op of S^1 in R^2}
        H =\frac{1}{a}\sigma_3 \qty(\sum_{i=1,2}\qty[\sigma_i\frac{\nabla_i-\nabla^\dagger_i}{2} +\frac{1}{2}\nabla_i \nabla^\dagger_i ]+\epsilon_A am ), 
    \end{align}
where we have introduced the Wilson term (the second term in the
square brackets) with the standard choice of the coefficient $w=1$. We assign the domain-wall mass by a step function
\begin{align}
    \epsilon_A(\hat{x},\hat{y})=\left\{ \begin{array}{cc}
        -1 & ((\hat{x},\hat{y})\in A)  \\
        1 & ((\hat{x},\hat{y})\notin A)
    \end{array}\right. .
\end{align}
The region $A$ inside the circle is defined by
\begin{align}
    A=\Set{(\hat{x},\hat{y})\in (\mathbb{Z}/N\mathbb{Z})^2 | \qty(\hat{x}-\frac{N-1}{2})^2+\qty(\hat{y}-\frac{N-1}{2})^2< (\hat{r}_0)^2},
\end{align}
where $\hat{r}_0=r_0/a$ is the radius of $S^1$ domain-wall in the lattice units and $(\frac{N-1}{2},\frac{N-1}{2})$ is the center of the circle. Note that $\frac{N-1}{2}$ needs not to be an integer. In fact, we take $N$ even for a reason explained below.

Since the domain-wall is curved, the chirality operator is
position-dependent. Let us translate $ \hat{x}-\frac{N-1}{2}\to \hat{x} $ and $\hat{y}-\frac{N-1}{2} \to \hat{y}$ so that the center of the circle is located at $(0,0)$, and $\hat{x}$ (resp. $\hat{y}$) takes a half integer when $N$ is even. Then, we can define the chirality operator on the lattice by
\begin{align}
    \gamma_{\text{normal}}:=& \sigma_1 \frac{\hat{x} }{\hat{r}}+\sigma_2 \frac{\hat{y}}{\hat{r}}
\end{align}
where $\hat{r}=\sqrt{ \hat{x}^2+\hat{y}^2}$. This operator is well-defined only when $N$ is even.

\begin{figure}[h]
  \begin{minipage}{0.45\linewidth}
    \centering
    \includegraphics[scale=0.5,bb= 0 0 461 346]{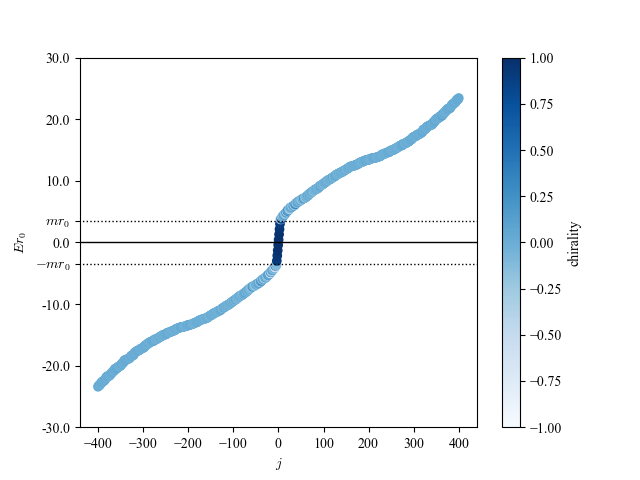}
  \end{minipage}
  \hfill
  \begin{minipage}{0.45\linewidth}
    \centering
    \includegraphics[scale=0.5,bb= 0 0 461 346]{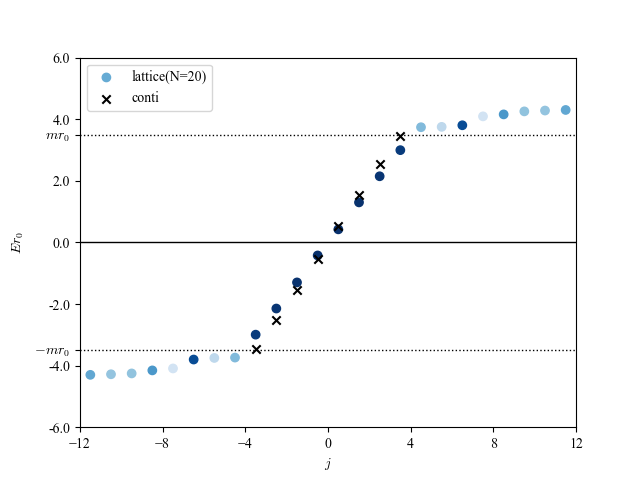}
  \end{minipage}
      \caption{The eigenvalue spectrum of the domain-wall Dirac operator at $ma = 0.7$, $\hat{r}_0=r_0/a = 5$ and $N = 20$. Circles denote the lattice data and crosses between $-mr_0$ and $mr_0$ represent the continuum counterpart. The gradation of the symbols represents its chirality. The left panel shows the whole spectrum and the right panel is a focus on the near-zero eigenvalues.
      }
    \label{fig:Eigenvalue S^1 lattice}
\end{figure}

We solve the eigenvalue problem of \eqref{eq:Hermitian Wilson Dirac op of S^1 in R^2} numerically and compute the expectation value of the chirality of each eigenstate. We plot the eigenvalues in Fig.~\ref{fig:Eigenvalue S^1 lattice} and represent their chirality by gradation of the symbol. Here we set the lattice size $N=20$, the radius of the circle domain-wall $\hat{r}_0=r_0 /a=5$, and the fermion mass $ma=0.7$. The circle symbols denote the lattice data and crosses represent their continuum limit. Here we label the eigenvalues with half integer $j$, expecting them to represent the eigenstates 
\begin{align}
    \cdots\leq E_{-\frac{3}{2}}\leq E_{-\frac{1}{2}} \leq 0 \leq E_{\frac{1}{2}} \leq E_{\frac{3}{2}} \leq \cdots.
\end{align}

\begin{figure}[]
 \begin{center}
  \subfigure{	
   \includegraphics[width=.45\columnwidth,bb=0 0 432 360]{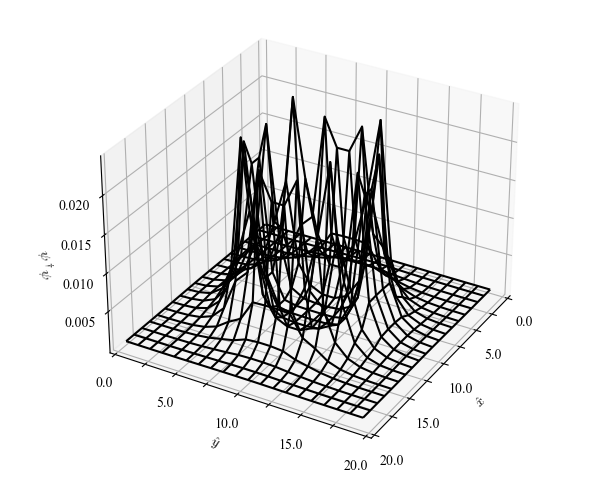}
  }\\ 
  \subfigure{
   \includegraphics[width=.45\columnwidth,bb=0 0 432 360]{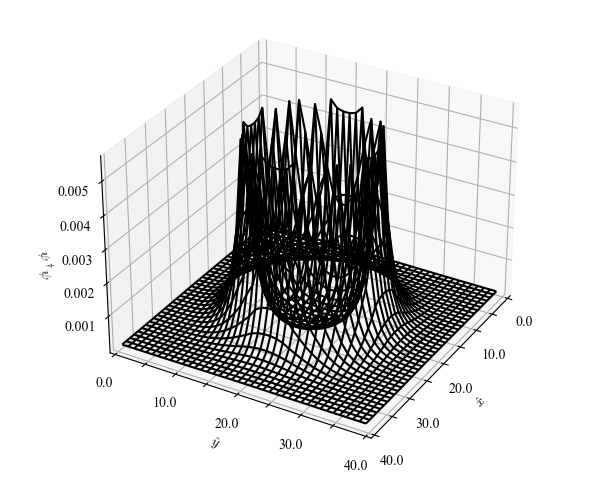}
  }~
  \subfigure{
   \includegraphics[width=.45\columnwidth,bb=0 0 432 360]{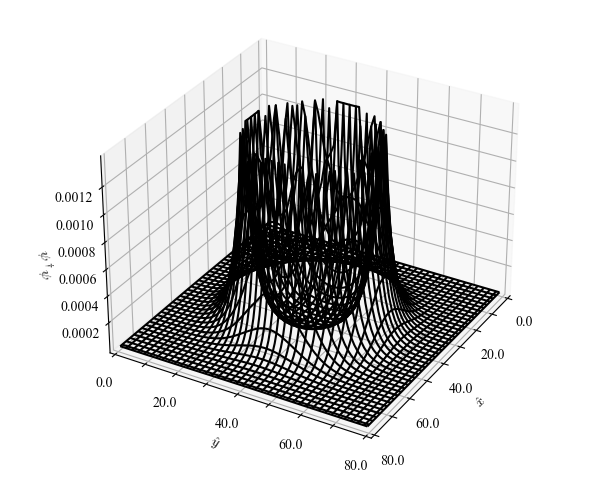}
  }
  \caption{Top panel: the amplitude of the eigenfunction with $E_{\frac{1}{2}}$ at the lattice spacing $a=1/N=L/20$. Bottom-left: the same plot as the top panel but with $a=L/40$. Bottom-right: the same but with $a=L/80$.} 
  \label{fig:S^1 edge state on lattice}
 \end{center}
\end{figure}

The eigenvalues of the near zero modes, whose absolute value is less than $m$ (indicated by the dotted lines), agree well with their continuum counterparts. The gap from zero, as the gravitational effect, is clearly seen. Moreover, they have the positive chirality as is expected. These states are localized at the curved domain-wall as shown in Fig.~\ref{fig:S^1 edge state on lattice} where the amplitude of the $E_\frac{1}{2}=0.4235/r_0$ eigenvector, which has the chirality $0.9902$ is plotted.

\begin{figure}[h]
  \begin{minipage}{0.45\linewidth}
    \centering
    \includegraphics[scale=0.5,bb=0 0 461 346]{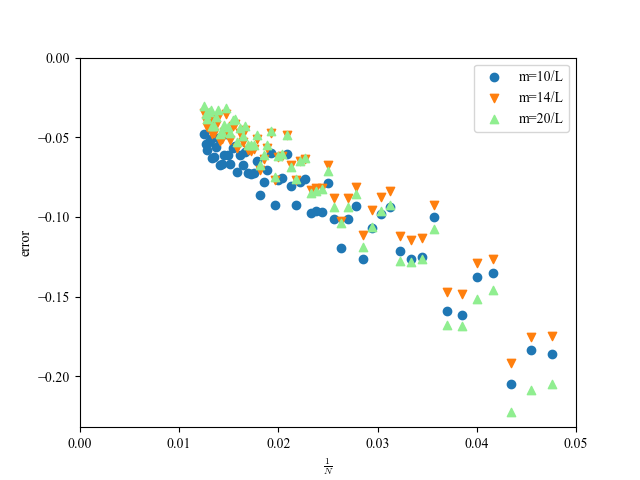}
  \end{minipage}
  \hfill
  \begin{minipage}{0.45\linewidth}
    \centering
    \includegraphics[scale=0.5,bb=0 0 461 346]{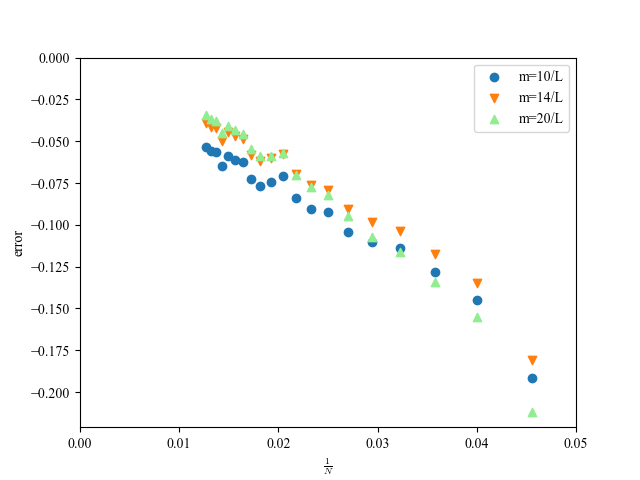}
  \end{minipage}
  \caption{Left panel: the relative deviation of the eigenvalue $\qty(E_{\frac{1}{2}}-E_\text{conti})/E_\text{conti}$, is plotted as a function of the lattice spacing $a=\frac{1}{N}$. Right: the same plot as the left panel but the three neighborhood points are averaged. 
  }
  \label{fig:continuum limit of error}
\end{figure}

\begin{figure}
    \centering
    \includegraphics[scale=0.8,bb=0 0 461 346]{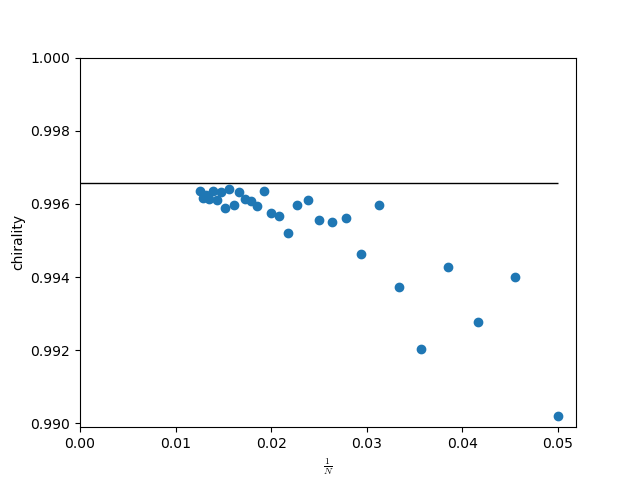}
    \caption{The chirality of the state with $E_\frac{1}{2}$ when $m=14/L$ and $\hat{r}_0= L/4$ is plotted as a function of the lattice spacing $a=\frac{1}{N}$. The horizontal line at $0.9966$ indicates the continuum value.
    }
    \label{fig:chirality}
\end{figure}

Let us discuss the systematics due to the finite lattice spacings. In Fig.~\ref{fig:continuum limit of error}, we plot the relative deviation of $E_{\frac{1}{2}}$ with three finite masses $m=10/L,14/L,20/L$ from the continuum result $E_\text{conti}$,
\begin{align}
    \text{error}=\qty(E-E_\text{conti})/E_\text{conti},
\end{align}
as a function of the lattice spacing $a=1/N$. Although some oscillation is visible in the left panel (which is tamed after three-point binning in the right panel), the data show a linear dependence on the lattice spacing $a$ to the continuum limit. We also find that the chirality expectation value approaches to the continuum value $0.9966$, as presented in Fig.~\ref{fig:chirality}. Note that the edge mode is not perfectly chiral even in the continuum theory, due to the finite value of $m$ and $r_0$.

\begin{figure}
    \centering
    \includegraphics[scale=0.8,bb=0 0 461 346]{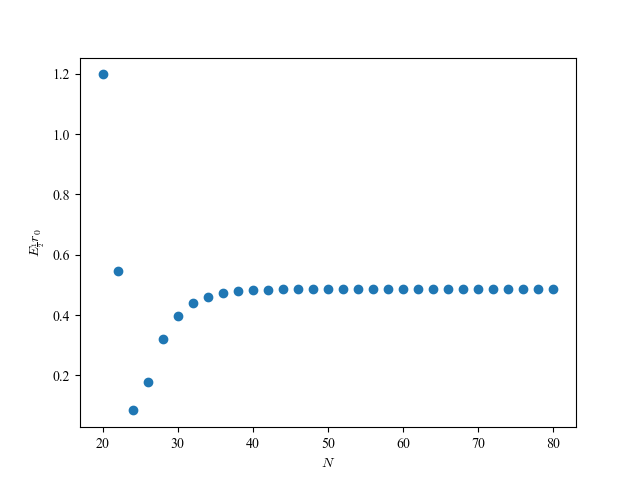}
    \caption{Finite lattice size $N$ scaling of the eigenvalue $E_{\frac{1}{2}}r_0$ at $ma=0.35$ and $\hat{r}_0=10$.
    }
    \label{fig:finite_volume_effect}
\end{figure}

Next, we discuss the finite-volume effects. In the numerical analysis on the lattice, we assign the periodic boundary condition in every direction with the finite size of $L$. In Fig.~\ref{fig:finite_volume_effect} we plot the eigenvalue $r_0E_{\frac{1}{2}}$ as a function of the lattice size $N=L/a$ with fixed values of $r_0=10a$ and $ma=0.35$. For the lattice size greater than $N=40$ or $L=4r_0$, the finite volume effect is negligible.




\begin{figure}
    \centering
    \includegraphics[scale=0.8,bb=0 0 461 346]{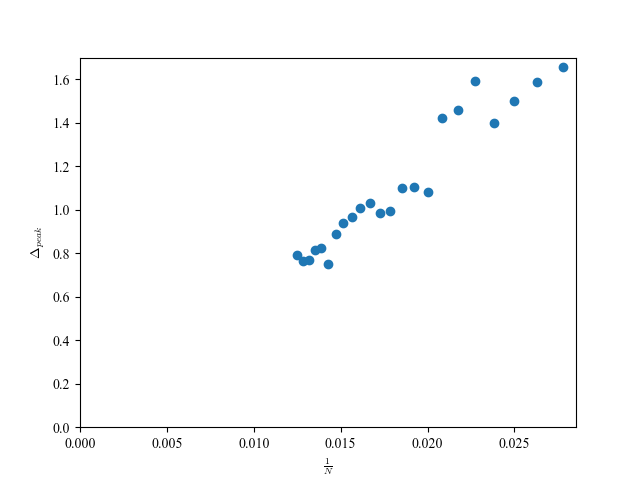}
    \caption{The rotational symmetry violation measured by the difference between the highest peak and the lowest peak of the amplitude of the eigenfunction with $E_\frac{1}{2}$ ($m=14/L$ and $r_0=L/4$). See the main text for the details.
    }
    \label{fig:recovery of rotational symm}
\end{figure}

Finally, let us address the recovery of the rotational symmetry. The zig-zag behavior of the peaks of the amplitude of the eigenfunction in Fig.~\ref{fig:S^1 edge state on lattice} reflects the violation of the rotational symmetry on the square lattice. However, we find that the spiky shapes become milder and milder when we decrease the lattice spacing. In order to quantify the rotational symmetry violation, we compare the highest and lowest peaks of the amplitude of the eigenfunction with $E_\frac{1}{2}$. The local amplitude $(\psi^\dagger \psi)_{(\hat{x},\hat{y}) }$ represents the probability distribution of the eigenstate on the $a \times a$ square at the lattice point $(\hat{x},\hat{y})$.  We collect a set of the peaks per $\hat{x}$ slices: 
\begin{align}
    P&= \Set{ \max_{\hat{y} }( \psi^\dagger\psi_{(\hat{x},\hat{y})} ) |  -\hat{r}_0<\hat{x}<\hat{r}_0}
\end{align}
and take the difference between the maximum and minimum of the set:
\begin{align}
    \Delta_{\text{peak}}=(\max(P)-\min(P))/a^2.
\end{align} 
We plot $\Delta_{\text{peak}}$ as a function of the lattice spacing $a=1/N$ in Fig.~\ref{fig:recovery of rotational symm}. Our data indicate automatic recovery of the rotational symmetry, as is naively expected from its recovery of the higher dimensional square lattice. 

\section{$S^2$ domain-wall in three-dimensional flat space
}\label{sec:S^2 in R^3}
In this section, we embed an $S^2$ into the three-dimensional Euclidean space as a domain-wall. Here, we locate its center at $(0,0,0)$ and denote its radius by $r_0$. In general, the massive Dirac operator $D+m$ is a complex operator in three dimensions. In order to make a well-defined eigenvalue problem, we introduce two flavors of the spinor fields. One can regard this system as a one-flavor four-dimensional fermion with one direction compactified to an infinitesimal size. In the latter interpretation, the existence of the chiral symmetry is obvious.

\subsection{Continuum analysis} 
In continuum theory, our target Hermitian Dirac operator is 
\begin{align}
    H= \gamma^5 \qty(\sum_{j=1}^3\gamma^j \pdv{}{x^j}+m\epsilon )=\mqty(m \epsilon & \sigma^j \partial_j \\ -\sigma^j \partial_j & -m\epsilon ) \label{eq:Hermitian Dira operator for S^2 in R^3},\ 
\end{align}
where the gamma matrices are given by a direct product of $2\times 2$ matrices: ${\gamma}^5=\sigma_3 \otimes 1,\ {\gamma}^j=\sigma_1\otimes \sigma_j$. The former $2\times 2$ matrices can be regarded as an operator on the two-flavor space. The mass parameter is denoted by $m$ and $\epsilon=\text{sign}(r-r_0)$ is a step function. Let us take the standard polar coordinate $(r,\theta, \phi)$ so that the radial direction is equal to the normal direction to the domain-wall, and the remaining $\theta$ and $\phi$ directions are tangent to it. This operator acts on the two flavors of the spinor fields in $\mathbb{R}^3$, or a four-component fermion.

First, let us solve the eigenproblem of \eqref{eq:Hermitian Dira
operator for S^2 in R^3} following the general recipe shown in Sec.
\ref{sec:Aconnection}. By a $Spin(3)\simeq SU(2)$ rotation,
\begin{align}
R(\theta,\phi)=\exp(\theta[\gamma^3,\gamma^1]/4)\exp(\phi[\gamma^1,\gamma^2]/4)=1\otimes\exp(i\theta\sigma_2/2)\exp(i\phi\sigma_3/2),
\end{align}
we can align the gamma matrices in the radial direction to $\gamma^3$ and that in the $\theta,~\phi$ directions to $\gamma^1$ and $\gamma^2$, respectively. 
However, this matrix is valid only locally and the same problem arises as in the previous section, in particular on the periodicity with respect to $\phi$. Therefore, we perform a further $U(1)$ rotation $\exp(-i\phi/2)$ to obtain the transformed spinor field $\psi^\prime=\exp(-i\phi/2)R(\theta,\phi)\psi$. Then the Hamiltonian is transformed as
\begin{align}
    H^\prime=&e^{-i\frac{\phi}{2}} R(\theta,\phi)   H R(\theta,\phi)^{-1} e^{i\frac{\phi}{2}} \nonumber \\
    =&\mqty( \epsilon m & \sigma_3 \qty( \pdv{}{r} +\frac{1}{r}+ \frac{1}{r}\sigma_3 \Slash{D}_{S^2}^\prime  )\\
    -\sigma^3 \qty( \pdv{}{r} +\frac{1}{r}+  \frac{1}{r} \sigma^3\Slash{D}_{S^2}^\prime ) & -\epsilon m)
\end{align}
and $\Slash{D}_{S^2}^\prime$ is 
\begin{align}\label{eq:Dirac op on S2}
      \Slash{D}_{S^2}^\prime=\qty(\sigma_1 \pdv{}{\theta} +\sigma_2 \qty( \frac{1}{\sin \theta} \pdv{}{\phi}+ \frac{i}{2\sin \theta} -\frac{\cos\theta}{2 \sin\theta} \sigma_1 \sigma_2 ) ) ,
\end{align}
where we find the nontrivial Spin and Spin$^c$ connections.

In a similar way to the previous section, the large $m$ limit requires the edge-localized solution to have the form
\begin{align}
    \psi^\prime
    =\frac{e^{-m\abs{r-r_0}} }{r}  \mqty( \chi^\prime(\theta,\phi) \\ \sigma_3 \chi^\prime( \theta,\phi)),
\end{align}
which has the positive ``chirality'' of  $\gamma^3= \mqty( 0 & \sigma_3 \\  \sigma_3 & 0)$,
or the gamma matrix in the normal direction in the original frame,
\begin{align}
        \gamma_{\text{normal}}:=&\sum_{i=1}^3 \frac{x^i}{r} {\gamma}^i.
    \end{align}
Here, $\chi$ converges to an eigenstate of the massless Dirac operator $\frac{1}{r}\Slash{D}^\prime_{S^2} \sigma_3 |_{r=r_0}= \frac{1}{r_0}\Slash{D}^\prime_{S^2} \sigma_3 $ on the $S^2$ domain-wall. As explicitly seen in \eqref{eq:Dirac op on S2}, the edge-localized modes feel gravity through the induced Spin and Spin$^c$ connections. 

According to \cite{Abrikosov2002Dirac}, $\Slash{D}_{S^2}^\prime \sigma_3$ is commutative with
\begin{align}
    J^\prime_{\pm}=\frac{e^{ \pm i\phi}}{\sqrt{2}} \qty{ \pm \pdv{}{\theta}+ i\frac{\cos \theta}{\sin \theta} \qty(\pdv{}{\phi}+\frac{i}{2})+\frac{1}{2\sin \theta} \sigma_3},~J_3^\prime=-i\pdv{}{\phi}+\frac{1}{2},
\end{align}
and a parity operator
\begin{align}
    P\chi(\theta,\phi)=\sigma_1 \chi(\pi-\theta,\phi+\pi). 
\end{align}
$J_\pm^\prime$ and $J_3^\prime$ satisfy
\begin{align}
    [J_+^\prime,J_-^\prime]=J^\prime_3,~[J^\prime_3,J^\prime_{\pm}]=\pm J_{\pm}^\prime.
\end{align}
Thus the eigenstate of $\Slash{D}_{S^2}^\prime \sigma_3 $ is that of $(J^\prime)^2=J_+^\prime J_-^\prime +J_-^\prime J_+^\prime+J_3^\prime J_3^\prime$, $J_3^\prime$ and $P$. Due to the nontrivial connections, the eigenvalues of $(J^\prime)^2$ and $J_3$ are represented by half-integers by $j(j+1)~(j=\frac{1}{2}, \frac{3}{2},\cdots)$, that of $J^\prime_3$ is denoted as $j_3=-j,-j+1,\cdots,j-1,j $. The highest eigenfunction with $(j,j_3=j)$ is obtained from the condition $J_+^\prime \chi_{j,j_3=j}^\prime =0$ as
\begin{align}
    \chi_{j,j_3=j,\pm }^\prime= (e^{i\phi} \sin \theta)^{j-\frac{1}{2}} \mqty(  \cos \frac{\theta}{2} \\ \mp \sin \frac{\theta}{2}),   
\end{align}
which has the eigenvalue $(-1)^{j\pm \frac{1}{2}}$ of $P$. By a direct substitution, we find that $\chi_{j,j_3=j,\pm }^\prime$ is the eigenstate of $\Slash{D}_{S^2}^\prime \sigma_3$ with the eigenvalue
\begin{align}\label{eq:eigenvalue of D_S2}
    \lambda=\pm \qty(j+\frac{1}{2}).
\end{align}
Since $J_-'$ commutes with $\Slash{D}_{S^2}^\prime \sigma_3$, every descendant state
\begin{align}
    \chi_{j,j_3,\pm}^\prime= (J_{-}^\prime)^{j-j_3} (e^{i\phi} \sin \theta)^{j-\frac{1}{2}} \mqty(  \cos \frac{\theta}{2} \\ \mp  \sin \frac{\theta}{2}),
\end{align}
shares the same eigenvalue \eqref{eq:eigenvalue of D_S2}. Namely, we have the $(2j+1)$-fold degeneracy.

Thus we find the edge-localized eigenmode as
    \begin{align}
    (\tilde{\psi}^{E}_{j,j_3,\pm})^\prime & \simeq \sqrt{\frac{m}{2}}\frac{e^{-m\abs{r-r_0}}}{r}\mqty(\chi_{j,j_3,\pm}^\prime \\ \sigma_3 \chi_{j,j_3,\pm}^\prime),\\
    E&\simeq \pm \frac{j+\frac{1}{2}}{r_0}.
\end{align}

The spectrum has a gap around $E=0$, which is a bigger footprint of the gravity than that seen in the $S^1$ domain-wall case. The absence of the zero eigenvalue is consistent with the vanishing theorem \cite{Friedrich1980DerersteEigenwert}, which states that the Dirac operator cannot have solutions on a manifold with non-negative curvature everywhere.

In fact, we can solve the Dirac equation in the original flat frame on $X=\mathbb{R}^3$ for finite $m$. We present the details in the appendix~\ref{App:direct}. We just present the results below. The edge-localized solutions are given by
\begin{align}
    \psi^{E>0}_{j,j_3,+}&=\left\{
\begin{array}{ll}
     \frac{A}{\sqrt{r}}\mqty( \sqrt{m^2-E^2} I_{j }(\sqrt{m^2-E^2}r) \chi_{j,j_3,+} \\(m+E) I_{j+1}(\sqrt{m^2-E^2}r) \frac{\sigma\cdot x}{r}\chi_{j,j_3,+} )  & (r<r_0) \\
     \frac{B}{\sqrt{r}}\mqty( (m+E) K_{j }(\sqrt{m^2-E^2}r)\chi_{j,j_3,+} \\\sqrt{m^2-E^2} K_{j+1}(\sqrt{m^2-E^2}r) \frac{\sigma\cdot x}{r}\chi_{j,j_3,+} ) & (r>r_0)
\end{array}
\right. ,    
\end{align}
and
\begin{align}
\psi^{E<0}_{j,j_3,-}&=\left\{
\begin{array}{ll}
\frac{A^\prime}{\sqrt{r}}\mqty( (m-E) I_{j+1}(\sqrt{m^2-E^2}r) \chi_{j,j_3,-} \\\sqrt{m^2-E^2} I_{j}(\sqrt{m^2-E^2}r) \frac{\sigma\cdot x}{r}\chi_{j,j_3,-} ) & (r<r_0) \\
     \frac{B^\prime}{\sqrt{r}}\mqty( \sqrt{m^2-E^2} K_{j+1 }(\sqrt{m^2-E^2}r) \chi_{j,j_3,-} \\(m-E) K_{j}(\sqrt{m^2-E^2}r) \frac{\sigma\cdot x}{r}\chi_{j,j_3,-} ) & (r>r_0)
\end{array}
\right. ,
\end{align}
where $A,~B,~A^\prime,~B^\prime$, and $E$ are determined by the continuity at $r=r_0$ and the normalization. The eigenvalue $E$ is determined as a solution to
\begin{align}
    \frac{I_{j}}{I_{j+1}}\frac{K_{j+1}}{K_{j}}(\sqrt{m^2-\abs{E}^2}r_0)=\frac{m+\abs{E}}{m-\abs{E}}.
\end{align}
It is a good exercise to confirm that the results are consistent with those in the $m\to \infty$ limit obtained in the rotated frame.

\subsection{Lattice analysis}
Let $(\mathbb{Z}/N \mathbb{Z})^3$ be a three-dimensional lattice space and $0\leq \hat{x},\hat{y},\hat{z} \leq N-1$ be the lattice coordinates. We impose the periodic boundary condition in every direction. We consider the Hermitian Wilson-Dirac operator as
    \begin{align}
        H =\frac{1}{a}\gamma^5 \qty(\sum_{i=1,2,3}\qty[\sum_{i=1}^3\gamma^i\frac{\nabla_i-\nabla^\dagger_i}{2} +\frac{1}{2}\nabla_i \nabla^\dagger_i ]+\epsilon_A am ), \label{eq:Hermitian Wilson Dirac op of S^2 in R^3}
    \end{align}
where the region $A$ inside the sphere is defined by
\begin{align}
        A=\Set{(\hat{x},\hat{y},\hat{z})\in (\mathbb{Z}/N\mathbb{Z})^3 | \qty(\hat{x}-\frac{N-1}{2})^2+\qty(\hat{y}-\frac{N-1}{2})^2+\qty(\hat{z}-\frac{N-1}{2})^2 < (\hat{r}_0)^2},
\end{align}
and 
\begin{align}
    \epsilon_A (\hat{x},\hat{y},\hat{z})= \left\{
    \begin{array}{cc}
        -1 & ( (\hat{x},\hat{y},\hat{z}) \in A) \\
        1 & ( (\hat{x},\hat{y},\hat{z}) \notin A)
    \end{array}
    \right.
\end{align}
is a step function which defines the $S^2$ domain-wall with the radius $r_0$. The center of the sphere is located at $(\frac{N-1}{2},\frac{N-1}{2},\frac{N-1}{2})$. Translating the center $(\frac{N-1}{2},\frac{N-1}{2},\frac{N-1}{2})$ to the origin, the chirality operator is defined by
\begin{align}\label{eq:chirality S2 in R3}
    \gamma_{\text{normal}}= \frac{\hat{x}}{\hat{r}} \gamma^1+ \frac{\hat{y}}{\hat{r}} \gamma^2 +\frac{\hat{z}}{\hat{r}} \gamma^3,
\end{align}
where $\hat{r}$ represents the length from the center of the circle. This operator is well-defined only when $N$ is even.

\begin{figure}[h]
  \begin{minipage}{0.45\linewidth}
    \centering
    \includegraphics[scale=0.5,bb=0 0 461 346]{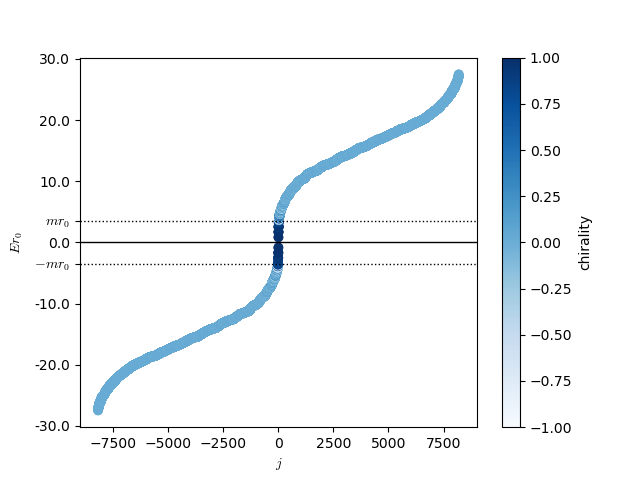}
  \end{minipage}
  \hfill
  \begin{minipage}{0.45\linewidth}
    \centering
    \includegraphics[scale=0.5,bb=0 0 461 346]{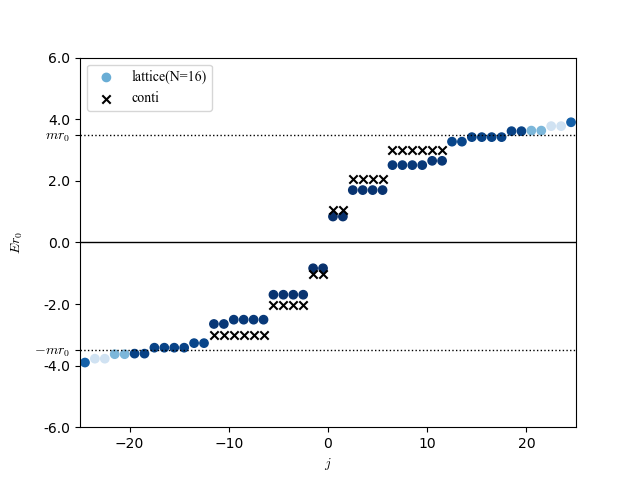}
  \end{minipage}
  \caption{The amplitude of the eigenfunction with $E_{\frac{1}{2}}$ at $ma=0.875$, $r_0/a =4$. In the left panel, the amplitude at every site in the whole three-dimensional lattice is represented by the gradation, while it is given by z-axis in the right panel focusing on the two-dimensional plane at $\hat{z}=7$.
  }
  \label{fig:eigenvalue of S2 in R3}
\end{figure}

We solve the eigenvalue problem of $H$ numerically. For the case with $ma=0.875$, $r_0/a =4$, and $N=16$, we plot the eigenvalue spectrum in Fig.~\ref{fig:eigenvalue of S2 in R3}, where we arrange the eigenvalues in ascending order of $j$ and the gradation represent their chirality.



\begin{figure}[h]
  \begin{minipage}{0.45\linewidth}
    \centering
    \includegraphics[scale=0.5,bb=0 0 461 346]{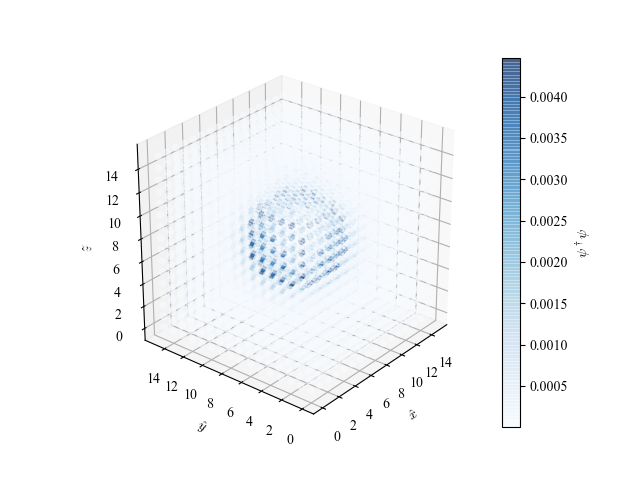}
  \end{minipage}
  \hfill
  \begin{minipage}{0.45\linewidth}
    \centering
    \includegraphics[scale=0.5,bb=0 0 461 346]{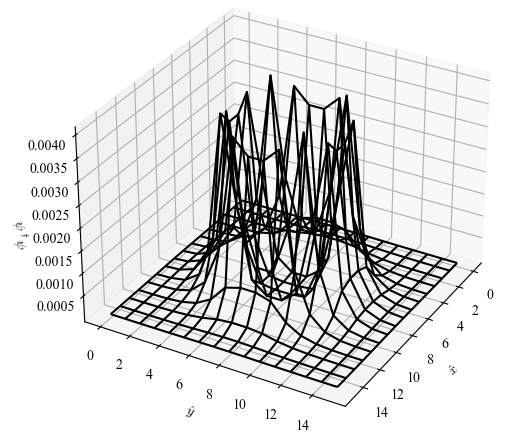}
  \end{minipage}
  \caption{The amplitude of the eigenfunction with $E_{\frac{1}{2}}$ at $ma=0.875$ and $r_0/a =4$. In the left panel, the amplitude at every site in the whole three-dimensional lattice is represented by the gradation, while it is given by $z$-axis in the right panel focusing on the two-dimensional plane at $\hat{z}=7$.
  }
  \label{fig:Eigenstate of S2 in R3}
\end{figure}

We can see a good agreement between the numerical lattice data (circle symbols) of $Er_0$ and the continuum results (crosses) between $-mr_0$ and $mr_0$, including the nontrivial degeneracy with respect to the different value of $j_3$. It is also consistent with the continuum prediction that the chirality is almost unity. The gap from zero indicates that these modes feel gravity through the induced Spin or Spin$^c$ connections. We notice that there are some chiral modes which do not have the continuum counterparts near $E=m$. We expect these modes will be absorbed into the bulk modes in the continuum limit when their continuum limit of the eigenvalues exceeds the mass. The near-zero modes are localized on the $S^2$ domain-wall as shown in Fig.~\ref{fig:Eigenstate of S2 in R3}, where the amplitude of the $E_\frac{1}{2}$ state is presented by the gradations.


\begin{figure}[h]
  \begin{minipage}{0.45\linewidth}
    \centering
    \includegraphics[scale=0.5,bb=0 0 461 346]{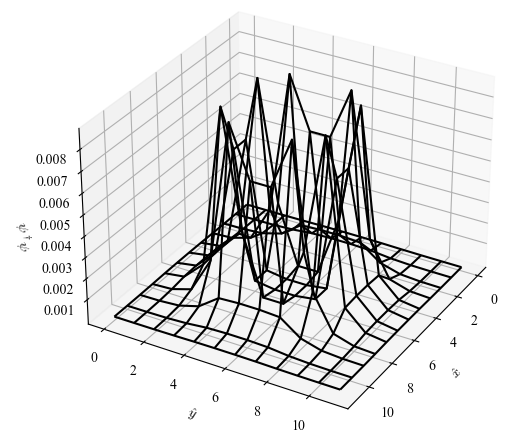}
  \end{minipage}
  \hfill
  \begin{minipage}{0.45\linewidth}
    \centering
    \includegraphics[scale=0.5,bb=0 0 461 346]{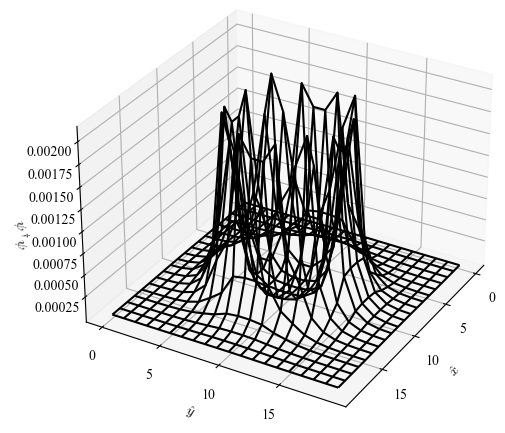}
  \end{minipage}
      \caption{The same plot as the right panel of Fig.~\ref{fig:Eigenstate of S2 in R3} but with different lattice spacings $a=L/12$ (left panel) and $a=L/20$ (right).
      }
    \label{fig:S2 eigenstates at N=12 and N=20}
\end{figure}

\begin{figure}
    \centering
    \includegraphics[scale=0.8,bb=0 0 461 346]{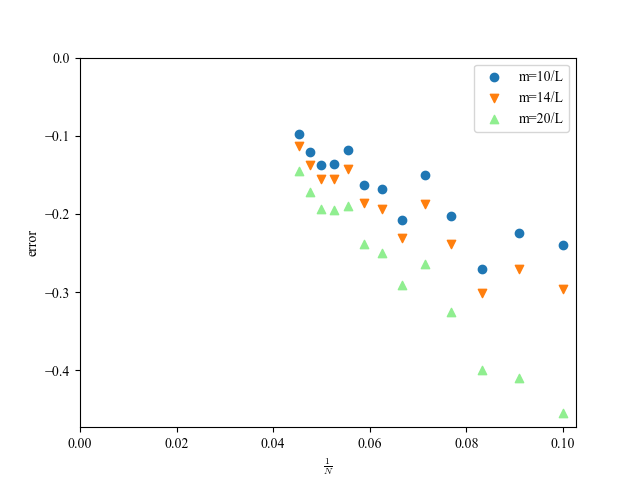}
    \caption{The relative deviation of the eigenvalue is plotted as a function of the lattice spacing $a=\frac{1}{N}$.}
    \label{fig:continuum limit of error S2}
\end{figure}

Let us discuss the systematics due to the lattice spacing. In Fig.~\ref{fig:continuum limit of error S2} we plot the deviation of $E_\frac{1}{2}$ with three masses $m=10/L,~14/L,~20/L$ and $r_0=L/4$ like the previous section. The data show a linear dependence on the lattice spacing $a$ to the continuum limit.

\begin{figure}
    \centering
    \includegraphics[scale=0.8,bb=0 0 461 346]{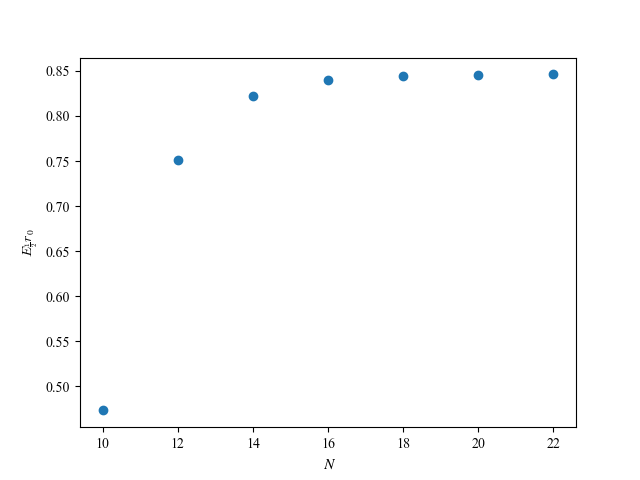}
    \caption{Finite lattice size $N$ scaling of the eigenvalue $E_{\frac{1}{2}}r_0$ at
    	$ma=0.875$ and $\hat{r}_0=4$.
    }
    \label{fig:S2 finite volume effect}
\end{figure}

Next, we discuss the finite volume effects. In Fig.~\ref{fig:S2 finite volume effect} we plot the eigenvalue $E_\frac{1}{2}r_0$ at $r_0 =4a$ and $ma=0.875$ in the same way as the previous section. A good convergence is seen after $N=L/a=16=4r_0/a$. 

\begin{figure}
    \centering
    \includegraphics[scale=0.8,bb=0 0 461 346]{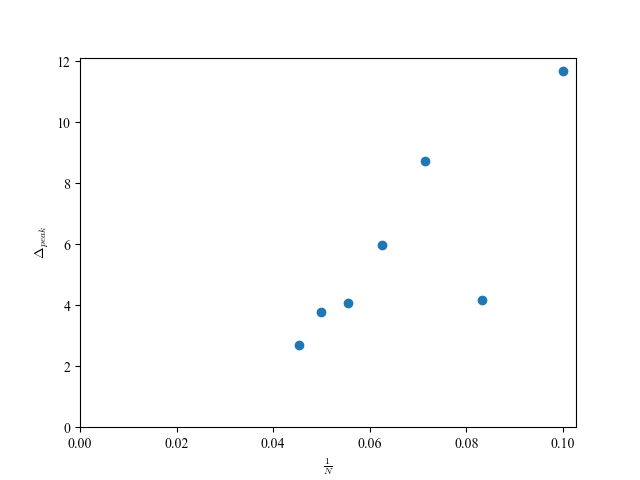}
    \caption{The rotational symmetry violation measured by the difference between the highest and lowest peaks in the $z$ direction of the amplitude of the eigenfunction with $E_\frac{1}{2}$ ($m=14/L$ and $r_0=L/4$) among different $x$ and $y$ points covering the spherical domain-wall. See the main text for the details.
    }
    \label{fig:recovery of rotational symm S2}
\end{figure}

The recovery of the rotational symmetry is also good. The smaller lattice spacing, the milder the spiky shape becomes in Fig.~\ref{fig:S2 eigenstates at N=12 and N=20}. To quantify the rotational symmetry violation, we define $\Delta_{\text{peak}}$ as the previous section but scanning a three-dimensional cube around the domain-wall. We plot the difference between the highest and lowest peaks in Fig.~\ref{fig:recovery of rotational symm S2}. Our data indicate automatic recovery of the rotational symmetry, as is naively expected from its recovery of the higher dimensional square lattice.

\section{Summary and Discussion}\label{sec:Conclusion}
In this work, we have investigated fermion systems on a square lattice having a curved domain-wall mass term. On the $S^1$ domain-wall embedded into $\mathbb{R}^2$ and that of $S^2$ into $\mathbb{R}^3$, we have shown that the edge-localized modes appearing at the domain-wall feel gravity through the induced Spin connection.

The effect of gravity or Spin connections is encoded in the spectrum of the domain-wall fermion Dirac operators. In particular, we have found a gap from zero in the eigenvalues of the Dirac operator, which is consistent with the vanishing theorem \cite{Friedrich1980DerersteEigenwert}, where the Dirac equation has no solution on a manifold with non-negative curvature everywhere.

For the cases of $S^1$ and $S^2$, we can analytically solve the continuum Dirac equation and compare the solutions with the lattice results. We have numerically solved the eigenproblem of the Hermitian lattice Dirac operator and found that the spectrum agrees well with that in the continuum theory. Our data at different lattice spacings indicates that the convergence to the continuum value is linear in $a$.

As was expected, we have found that the near-zero eigenmodes of the Dirac operator are localized at the curved domain-walls. Although they have spiky shapes due to the rotational symmetry breaking, they monotonically become flat in the naive continuum extrapolation. This observation suggests that the rotational symmetry on the spheres is automatically recovered together with the simple classical continuum limit.

Our numerical analysis has been done on a lattice having periodic boundary conditions. The large volume extrapolation of the data indicates a good saturation when the lattice size is four times larger than the radius of the spherical domain-walls. The finite systematic is thus controlled in the same manner as the conventional flat domain-wall fermion.


In this work, we have fixed the shape and location of the domain-walls. It is interesting to ask what happens when the domain-wall is allowed to move. On the lattice, the configuration of the domain-wall can be given by assigning the $\pm 1$ values to each lattice site. If we define the path-integral of these $\mathbb{Z}_2$ site valuables, in such a way that its dynamics depend only on the effective curvature of the domain-walls, it may give a novel realization of quantum gravity coupled to the edge-localized chiral fermions.


In the examples given in this paper, we did not have zero eigenmodes of the effective Dirac operator with nonzero gravitational potentials. In these cases, the Atiyah-Singer index is trivially zero. In recent studies of the domain-wall fermions \cite{Fukaya_2017Atiyah-Patodi-Singer,fukayaFuruta2020physicistfriendly,FukayaFurutaMatsuki2021Aphysicist-friendly,FukayaKawai2020TheAPS}, the nontrivial Atiyah-Patodi-Singer index can be realized to describe the bulk and edge correspondence of anomaly inflow. Then a natural question is if we can describe the gravitational anomaly inflow with the curved domain-wall fermions, and describe the related index theorem on the lattice. Mixed anomaly with the gauge link variables and the gravity will be also interesting.

\section*{Acknowledgment}
We thank M. Furuta, K. Hashimoto, M. Kawahira, S. Matsuo, T. Onogi, S. Yamaguchi and M. Yamashita for useful discussions. This work was supported in part by JSPS KAKENHI Grant Number JP18H01216 and JP18H04484.

\bibliographystyle{ptephy}
\bibliography{ref}

\appendix

\section{Weyl fermions on the domain-wall}\label{App:Weyl}

In the literature both in continuum theory \cite{Jackiw1976Solitons,CALLAN1985427Anomalies} and on a lattice \cite{KAPLAN1992342AMethod,Shamir1993Chiral}, the domain-wall fermion in odd dimensions was employed to describe Weyl fermions. The Dirac operator in that case is
\begin{align}
    D=\Slash{D}+\epsilon m
\end{align}
which is not Hermitian. In the main text of this paper, we introduce two-flavor degrees of freedom to make a Hermitian Dirac operator. In this appendix, let us consider the single flavor case.

Taking the frame decomposing the normal and tangent directions to the domain-wall, and the local scale transformation of the spinor field on $X=\mathbb{R}^n$: $\psi= \qty(g^{IJ}\pdv{f}{x^I}\pdv{f}{x^J})^{+\frac{1}{4}} \psi^\prime $, the operator acts on $\psi^\prime$ as
\begin{align}
    D^\prime
    =&  \underbrace{\gamma^a \qty(e_a +\frac{1}{4}\sum_{bc}\omega_{bc,a} \gamma^b\gamma^c)}_{\tilde{\Slash{D}}} +\gamma^{n+1}\pdv{}{t}+F +\epsilon m.
\end{align}

In the large $m$ limit, the near zero modes must converge to
the edge-localized form
\begin{align}
    \psi^\prime = e^{-m\abs{t}} e^{-\int_0^t dt^\prime F(y,t^\prime)}
\chi_+, 
\end{align}
where $\chi_+$ has the positive chirality of $\gamma^{n+1}$. The opposite chiral state appears for $D'^\dagger$. Thus, the Dirac operator $D'$ and $D'^\dagger$ effectively act as $\tilde{\Slash{D}}_\pm =\tilde{\Slash{D}}\frac{1}{2} (1 \pm \gamma^{n+1})$, respectively, on the chiral edge modes. Thus, we have Weyl fermions as the edge modes with the nontrivial Spin connection induced by the curved domain-wall.

\section{The direct computation of the edge modes} \label{App:direct}
In this appendix, we solve the eigenproblem without taking the large $m$ limit in the original frame of $X=\mathbb{R}^3$.
$H$ is commutative with the following three operators:
\begin{align}
    J_i&=1\otimes \hat{J}_i=1\otimes (L_i+\frac{1}{2}\sigma_i),\\
    J^2&=J_1^2 +J_2^2+J_3^2 =1\otimes \hat{J}^2\\
    P\psi(x)&=(\sigma_3\otimes 1) \psi(-x),
\end{align}
where $L_i=-i \epsilon_{ijk} x^j \partial_k$ is an orbital angular momentum operator, $\hat{J}_i$ denotes the total angular-momentum, and $P$ is a parity operator. Therefore, let us label our two-component spinor $\chi_{j,j_3,\pm}$ by the eigenvalues of $ J^2, J_3,P$:
\begin{align}
    \hat{J}^2\chi_{j,j_3,\pm}&=j(j+1)\chi_{j,j_3,\pm}\\
    \hat{J}_3  \chi_{j,j_3,\pm}&= j_3 \chi_{j,j_3,\pm} \\
    \chi_{j,j_3,\pm}(-x)&=(-1)^{j\mp \frac{1}{2}}\chi_{j,j_3,\pm}(x) \\
    \chi_{j,j_3,-}&=\frac{\sigma\cdot x}{r}\chi_{j,j_3,+}.
\end{align}
Note that we can write them explicitly using the spherical harmonics. 

We obtain the eigenstate with energy $E>0$ as
\begin{align}\label{eq:edgestate S^2 E>0}
    \psi^{E>0}_{j,j_3,+}&=\left\{
\begin{array}{ll}
     \frac{A}{\sqrt{r}}\mqty( \sqrt{m^2-E^2} I_{j }(\sqrt{m^2-E^2}r) \chi_{j,j_3,+} \\(m+E) I_{j+1}(\sqrt{m^2-E^2}r) \frac{\sigma\cdot x}{r}\chi_{j,j_3,+} )  & (r<r_0) \\
     \frac{B}{\sqrt{r}}\mqty( (m+E) K_{j }(\sqrt{m^2-E^2}r)\chi_{j,j_3,+} \\\sqrt{m^2-E^2} K_{j+1}(\sqrt{m^2-E^2}r) \frac{\sigma\cdot x}{r}\chi_{j,j_3,+} ) & (r>r_0)
\end{array}
\right.
\end{align}
and eigenstates with energy $E<0 $ as
\begin{align}\label{eq:edgestate S^2 E<0}
    \psi^{E<0}_{j,j_3,-}&=\left\{
\begin{array}{ll}
\frac{A^\prime}{\sqrt{r}}\mqty( (m-E) I_{j+1}(\sqrt{m^2-E^2}r) \chi_{j,j_3,-} \\\sqrt{m^2-E^2} I_{j}(\sqrt{m^2-E^2}r) \frac{\sigma\cdot x}{r}\chi_{j,j_3,-} ) & (r<r_0) \\
     \frac{B^\prime}{\sqrt{r}}\mqty( \sqrt{m^2-E^2} K_{j+1 }(\sqrt{m^2-E^2}r) \chi_{j,j_3,-} \\(m-E) K_{j}(\sqrt{m^2-E^2}r) \frac{\sigma\cdot x}{r}\chi_{j,j_3,-} ) & (r>r_0)
\end{array}
\right. ,
\end{align}
where $A,~B$ and $A^\prime,~B^\prime$ are constants and determined by the continuity at $r=r_0$ and the normalization. The eigenvalue $E$ satisfies a solution to
\begin{align}\label{eq:condition of E S^2}
    \frac{I_{j}}{I_{j+1}}\frac{K_{j+1}}{K_{j}}(\sqrt{m^2-\abs{E}^2}r_0)=\frac{m+\abs{E}}{m-\abs{E}}.
\end{align}
Note that the sign of $E$ is corresponding to the eigenvalue of the parity operator $P$.

In the large mass limit or $m\gg E$, the energy converges to 
\begin{align}
    E\simeq \pm \frac{j+\frac{1}{2}}{r_0},\ \qty(j=\frac{1}{2}, \frac{3}{2}\cdots),
\end{align}
where $\pm$ is corresponding to the eigenvalue $(-1)^{j \mp \frac{1}{2}}$ of $P$. We can see the gap from zero as a gravitational effect. 

The normalized eigenstate in that limit is obtained as
\begin{align}
    \tilde{\psi}^{E}_{j,j_3,\pm}\simeq \sqrt{\frac{m}{2}}\frac{e^{-m\abs{r-r_0}}}{r}\mqty(\chi_{j,j_3,\pm} \\ \frac{\sigma \cdot x}{r}\chi_{j,j_3,\pm}), 
\end{align}
which is an eigenstates of the gamma matrix 
\begin{align}
        \gamma_{\text{normal}}:=&\sum_{i=1}^3 \frac{x^i}{r} {\gamma}^i
    \end{align}
facing to the normal direction of $S^2$ and it has an eigenvalue $+1$.

\end{document}